# Strong atom-field coupling for Bose-Einstein condensates in an optical cavity on a chip

Yves Colombe[1,*], Tilo Steinmetz[1,2,*], Guilhem Dubois[1], Felix Linke[1,†], David Hunger[2] & Jakob Reichel[1]

*[1]Laboratoire Kastler Brossel, ENS / UPMC-Paris 6 / CNRS, 24 rue Lhomond, 75005 Paris, France. [2]Max-Planck-Institut für Quantenoptik / LMU, Schellingstr. 4, 80799 München, Germany. [†]Present address: BMW Group, Abt. Instrumentierung und Displays, Knorrstr. 147, D-80788 München. [*]These authors contributed equally to this work.*

**An optical cavity enhances the interaction between atoms and light, and the rate of coherent atom-photon coupling can be made larger than all decoherence rates of the system. For single atoms, this "strong coupling regime" of cavity quantum electrodynamics[1,2] (cQED) has been the subject of spectacular experimental advances, and great efforts have been made to control the coupling rate by trapping[3,4] and cooling the atom[5,6] towards the motional ground state, which has been achieved in one dimension so far[5]. For N atoms, the three-dimensional ground state of motion is routinely achieved in atomic Bose-Einstein condensates (BECs)[7], but although first experiments combining BECs and optical cavities have been reported recently[8,9], coupling BECs to strong-coupling cavities has remained an elusive goal. Here we report such an experiment, which is made possible by combining a new type of fibre-based cavity[10] with atom chip technology[11]. This allows single-atom cQED experiments with a simplified setup and realizes the new situation of N atoms in a cavity each of which is identically and strongly coupled to the cavity mode[12]. Moreover, the BEC can be positioned deterministically anywhere within the cavity and localized entirely within a single antinode of the standing-wave cavity field. This gives rise to a controlled, tunable coupling rate, as we confirm experimentally. We study the heating rate caused by a cavity transmission measurement as a function of the coupling rate and find no measurable heating for strongly coupled BECs. The spectrum of the coupled atoms-cavity system, which we map out over a wide range of atom numbers and cavity-atom detunings, shows vacuum Rabi splittings exceeding 20 gigahertz, as well as an unpredicted additional splitting which we attribute to the atomic hyperfine structure. The system is suitable as a light-matter quantum interface for quantum information[13].**

The interaction of an ensemble of $N$ atoms with a single mode of radiation has been a recurrent theme in quantum optics at least since the work of Dicke[14], who showed that under certain conditions the atoms interact with the radiation collectively, giving rise to new effects such as superradiance. Recently, collective interactions with weak fields, with and without a cavity, have become a focus of theoretical and experimental investigations, especially since it became clear that they can turn the ensemble into a quantum memory[13,15]. Such a memory would become a key element for processing quantum information[13,16] if realized with near-unit conversion efficiency and long storage time. The figure of merit determining the probability of converting an atomic excitation into a cavity photon (a "memory qubit" into a "flying qubit") is the collective cooperativity $C_N = \frac{g_N{}^2}{2\kappa\gamma}$, where $g_N$ is the collective coupling strength[17] between the ensemble and the field, $2\kappa$ is the cavity photon decay rate and $2\gamma$ the atomic spontaneous emission rate. (Up to a factor of order 1, $C_N$ is the single-pass optical depth of the atomic sample multiplied by the cavity finesse $\mathcal{F}$.) For weak excitation, to which we restrict ourselves throughout this article, $g_N = \sqrt{N}\bar{g}_1$, where $\bar{g}_1{}^2 = \int \frac{\rho(\mathbf{r})}{N} |g_1(\mathbf{r})|^2 \, d\mathbf{r}$, $g_1(\mathbf{r})$ is the position-dependent single-atom coupling strength and $\rho(\mathbf{r})$ is the atomic density distribution: the ensemble couples to the mode as a single "superatom" with a coupling strength increased by $\sqrt{N}$. In the strong-coupling regime $g_N > \kappa, \gamma$, which is realized in our cavity even for $N$=1, the atomic ensemble oscillates between its ground state and a symmetric excited state where a single excitation is shared by all the atoms. The coupled atoms – cavity system has dressed states of energies

$$E_\pm = \hbar\omega_A + \frac{\hbar}{2}\left(\Delta_C \pm \sqrt{\Delta_C{}^2 + 4g_N{}^2}\right), \qquad (1)$$

where $\Delta_C = \omega_C - \omega_A$, $\omega_C$ and $\omega_A$ being the cavity and atom resonance frequencies. When the cavity is tuned to atomic resonance ($\Delta_C$=0), these states are separated by the vacuum Rabi frequency[1,2] $2g_N$. The collective interaction and $\sqrt{N}$ scaling are not a consequence of atomic quantum statistics, but apply to thermal and quantum degenerate bosons and fermions, and even to nonidentical particles having the same transition moment[2,18], for a wide range of conditions in which interparticle correlations in the initial atomic state are negligible (see Supplementary Notes). Nevertheless, the spatial coherence, fundamentally lowest kinetic energy and smallest size of a BEC influence its interaction with light and make it the most desirable atomic state in many situations of conceptual and practical interest: the BEC allows to maximize the coupling and avoid decoherence effects associated with spatial inhomogeneities and with atomic motion[15]. In the free-space case, some of these aspects have been shown in detail for the case of superradiance, where reduced Doppler broadening in the condensate increased the coherence time by a factor of 30 over a thermal cloud at the transition temperature, making the effect observable only in the BEC[19]. Here we take advantage of the fact that the BEC has the smallest possible position spread in a given trap. This allows us to load the BEC into a single site of a far-detuned intracavity optical lattice. By choosing the lattice site, we achieve well-defined, maximised atom-field coupling in the standing-wave cavity field, where the local coupling varies as $g_1(x,r_\perp)=g_0\cos(2\pi x/\lambda_C)\exp(-r_\perp^2/w^2)$ ($x$ and $r_\perp$ are the longitudinal and transverse atomic coordinates, $w$ and $\lambda_C=2\pi c/\omega_C$ are the mode radius and wavelength). This is an important improvement for applications such as the quantum memory. Furthermore, BEC-cQED experiments such as ours and a simultaneous similar one[12] can be used to study BECs in the regime of very small atom numbers where the mean-field approximation breaks down, and may allow observation of effects such as a predicted slight modification of the refractive index of the atomic sample close to the BEC transition[20], and differences between quantum phases in transmission spectra of cavities containing a degenerate gas in an optical lattice[21].

We have developed a novel type of fibre-based Fabry-Perot (FFP) cavity[10,22] which achieves large single-atom peak coupling rates $g_0$ through reduced mode volume and high mirror curvature (see Methods), without the difficulties associated with evanescent fields in microtoroidal[23] or microsphere cavities. The setup is shown in figure 1. The combination ($g_0=2\pi\times215$MHz, $\kappa=2\pi\times53$MHz, $\gamma=2\pi\times3$MHz) places our cavity in the single-atom strong coupling regime and leads to a high single-atom cooperativity $C_0=145$. Despite its finesse which is an order of magnitude below standard cQED cavities, performance data of our cavity are comparable or superior to most of those, while all the dynamics occurs on a faster timescale. Two laser beams are coupled into the cavity through the input fibre, a weak tunable probe beam (frequency $\omega_L$), and an optional far-detuned beam at $\lambda_D=830.6$nm used to form a one-dimensional optical lattice along the cavity axis. Using a combination of chip currents and external magnetic fields, a BEC or cold thermal cloud of $^{87}$Rb atoms in the $|F=2,m_F=2\rangle$ ground state is prepared inside the cavity, and then positioned anywhere within the cavity mode. The slow axis of the magnetic trap can be oriented along the cavity axis $x$ or perpendicular to it along $y$. Atom-field interaction is studied either directly in this trap, or after switching on the far-detuned lattice to increase the confinement and gain control over the coupling rate. For magnetic trapping and also for weak lattices, we are able to observe an intact BEC after its interaction with the cavity field (Fig. 1d). For lattices with trapping frequency above $\nu_x\approx20$kHz, technical fluctuations in the lattice potential[3] heat up the condensate.

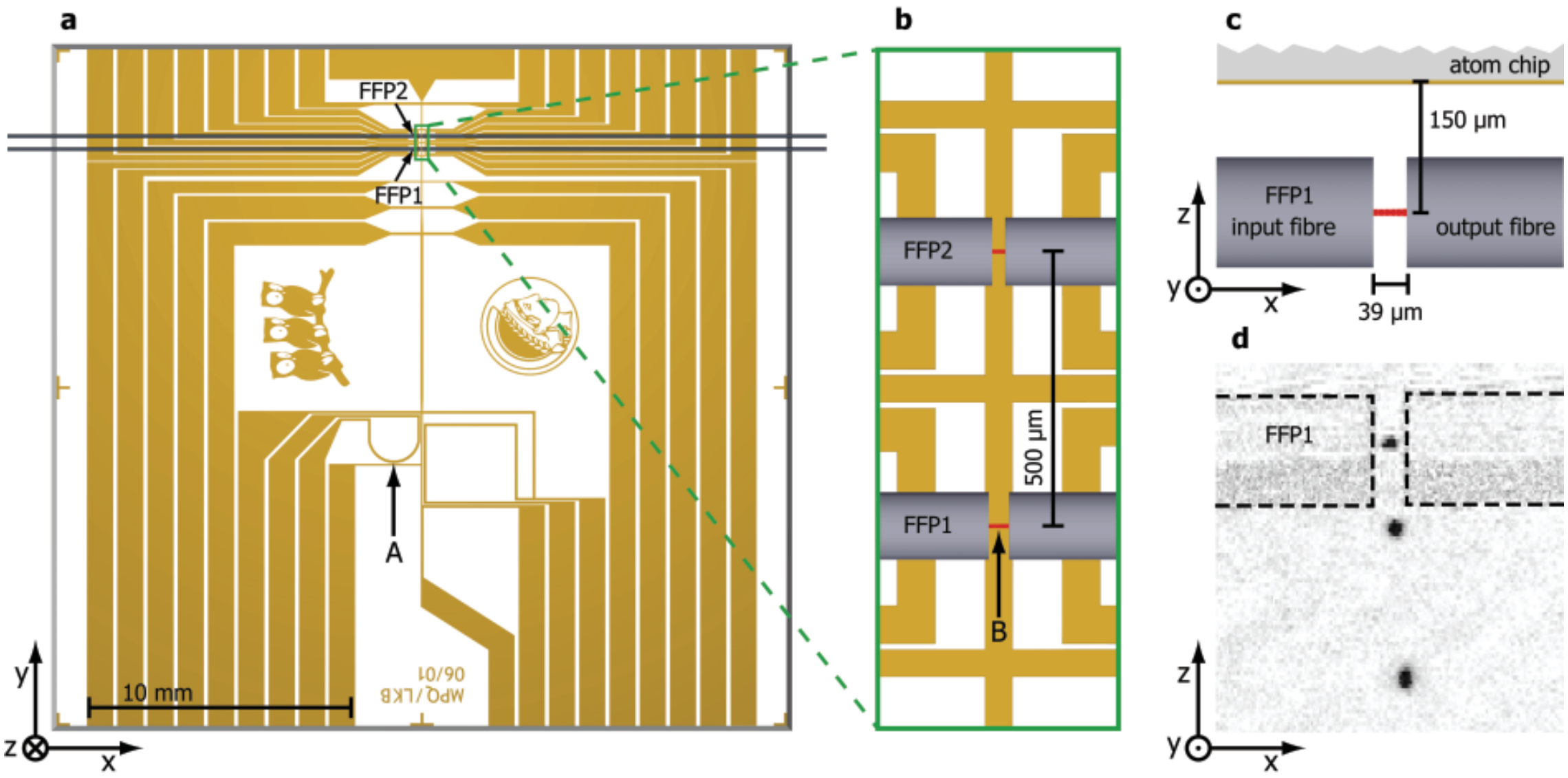

Figure 1: **Experimental setup**. **a**, Layout of the atom chip. "A" indicates the location of the first magnetic trap, loaded from a magneto-optical trap. **b**, Close-up view of the two fibre-Fabry-Perot (FFPs) optical cavities that are mounted on the chip. Cavity modes are drawn to scale in red. The BEC is produced in a magnetic trap and positioned in the FFP1 mode ("B"). **c**, Geometry of the FFP1 cavity. **d**, Overlay of three CCD time-of-flight (TOF) absorption images showing the anisotropic expansion of a BEC having interacted during 50ms with the cavity field under conditions similar to Fig. 4. The optical fibres are outlined for clarity.

In the following, we describe three experiments that explore the main aspects of atoms-field interaction in our system. First, we study the position dependence of $g_N(x)$ and show that a full control is achieved. Second, we observe the dependence of $g_N$ on atom number and map out the energies of the dressed states. Finally, we investigate the heating of a condensate by the intracavity field.

To study the position dependence of $g_N(x)$, we start by placing a BEC containing $N\approx1000$ atoms at a position $x_a$ on the cavity axis in a magnetic trap oriented along $y$ ($\nu_{x,z}$=2.7kHz, $\nu_y$=230Hz, bias field $B_y\approx1$G). We then ramp up a tight optical lattice with trapping frequencies $\nu_x$=50kHz, $\nu_{y,z}$=2.4kHz. The loaded atoms are now strongly confined in the combined trap, even though no longer Bose-condensed due to the technical heating. As the lattice and the probed cavity mode have different wavelengths $\lambda_D$ and $\lambda_L$, their overlap is modulated with a period $\lambda_D\lambda_L/2(\lambda_D-\lambda_L)$=6.4µm (Fig. 2a). We measure $g_N(x_a)$ by sweeping the probe laser detuning $\Delta_L=\omega_L-\omega_A=2\pi\times(+1\ldots-13)$GHz in 50ms, with cavity detuning $\Delta_C=\omega_C-\omega_A=0$ A transmission peak occurs at $|\Delta_L|=g_N$ when the lower dressed state is excited. Figure 2b shows the result for $x_a$ values spanning the full cavity length. We are able to reproduce the observed $g_N(x_a)$ by calculating the coupling of a Gaussian cloud centred on a single lattice site. The corresponding fit (red line on Fig. 2b) using the cloud diameter $2\sigma_x$ as a free parameter gives the value $2\sigma_x$=130nm, from which we deduce $2\sigma_y$=2.7μm, $2\sigma_z$=1.8μm based on the known ratio of the trapping frequencies and assuming thermal equilibrium ($T$=4.4μK). This fit gives a good indication of $\sigma_x$, but does not prove single-site loading because a similar fit is obtained when considering several clouds in adjacent sites, each with $2\sigma_x$=130nm.

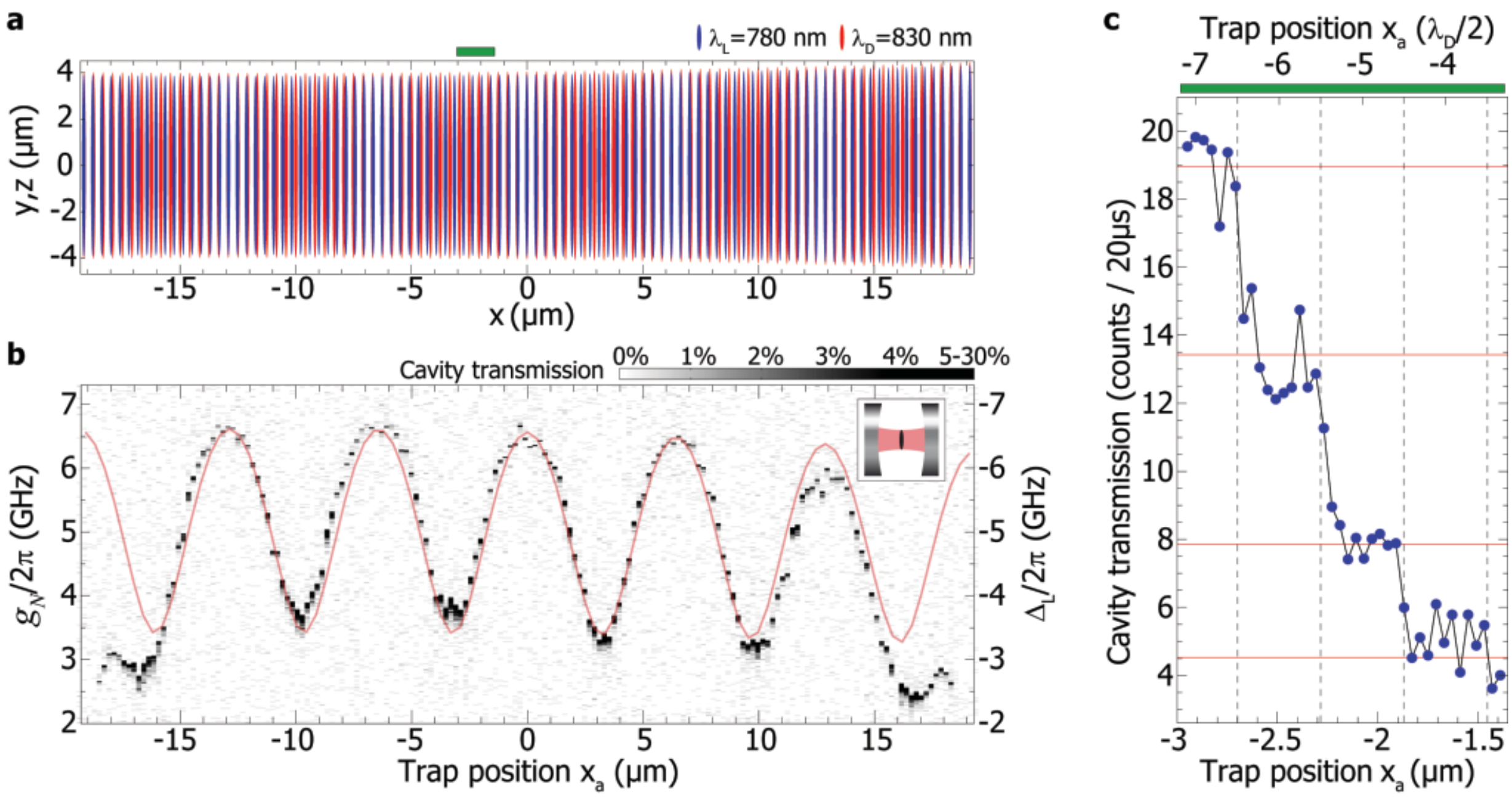

Figure 2: **Control of the coupling along resonator axis**. The BEC is brought to a position $x_a$ on the cavity axis and loaded into the optical lattice. **a**, The probe laser ($\lambda_L$=780.2nm) and optical lattice ($\lambda_D$=830.6nm) standing waves in the cavity have a variable overlap with 6.4μm period. The green bar indicates the region of the measurement in c. **b**, The loaded atoms show a strongly modulated coupling $3.4\text{GHz} \leq g_N(x_a)/2\pi \leq 6.6\text{GHz}$, depending on the local overlap between lattice and probe field. The rapid decrease of $g_N$ at the extremities of the mode is probably due to atom loss caused by collisions with the mirrors. The pictogram inset shows the orientation of the magnetic trap relative to the cavity mode. Probe intensity corresponds to a mean intracavity photon number for resonant cavity without atoms of $n_{res} \approx 5.5\times10^{-2}$. (Here and in the following experiments, we have checked that intensity-dependent peak shifts are negligible.) Maximum transmission in the figure corresponds to an intracavity photon number $n \approx 1.1\times10^{-2}$. **c**, The transmission of the cavity, probed at $\Delta_L=\Delta_C=2\pi\times$-100GHz with position increments $\delta x_a$=40nm, exhibits well-separated steps owing to the loading into successive single lattice sites. Vertical lines indicate the intensity minima of the optical lattice. Horizontal lines are expected transmission values with atoms localized in successive single lattice sites. Probe intensity: $n_{res} \approx 3.3\times10^{-2}$.

To demonstrate unambiguously the transfer into a single lattice site, we measure the transmission as a function of $x_a$ in the dispersive regime, $\Delta_L=\Delta_C=2\pi\times$-100GHz, where it depends on the atoms-induced cavity resonance shift $\delta\omega_C=g_N^2/\Delta_L$. If and only if single-site loading is achieved, transmission should change in steps between $x_a$ values corresponding to adjacent lattice sites. We use reduced increments $\delta x_a$=40nm, a BEC with $N\approx600$ and a magnetic trap with $\nu_{x,z}$=4kHz, $\nu_y$=230Hz. Using analytical formulas[24] for the 1d-3d crossover regime applying to this BEC we find a central radial diameter $2\sigma_{x,z}=2\times1.37\sigma_{ho}$=330nm, where $\sigma_{ho}$ is the harmonic-oscillator ground state radius. This BEC is loaded into a lattice with $\nu_x$=100kHz, $\nu_{y,z}$=4.8kHz (destroying the condensate, but producing stronger confinement along $x$), and the magnetic potential switched off. Figure 2c shows the transmission of the cavity. Each point is averaged over two experimental runs. Well-separated plateaus are observed, corresponding to discrete values of $g_N$ in good agreement with the calculated transmissions for ensembles localized in a single lattice site with diameter $2\sigma_x$=100nm (horizontal lines). This experiment shows the deterministic transfer of the atom cloud into successively addressed single sites of the lattice, each of which is differently coupled to the cavity.

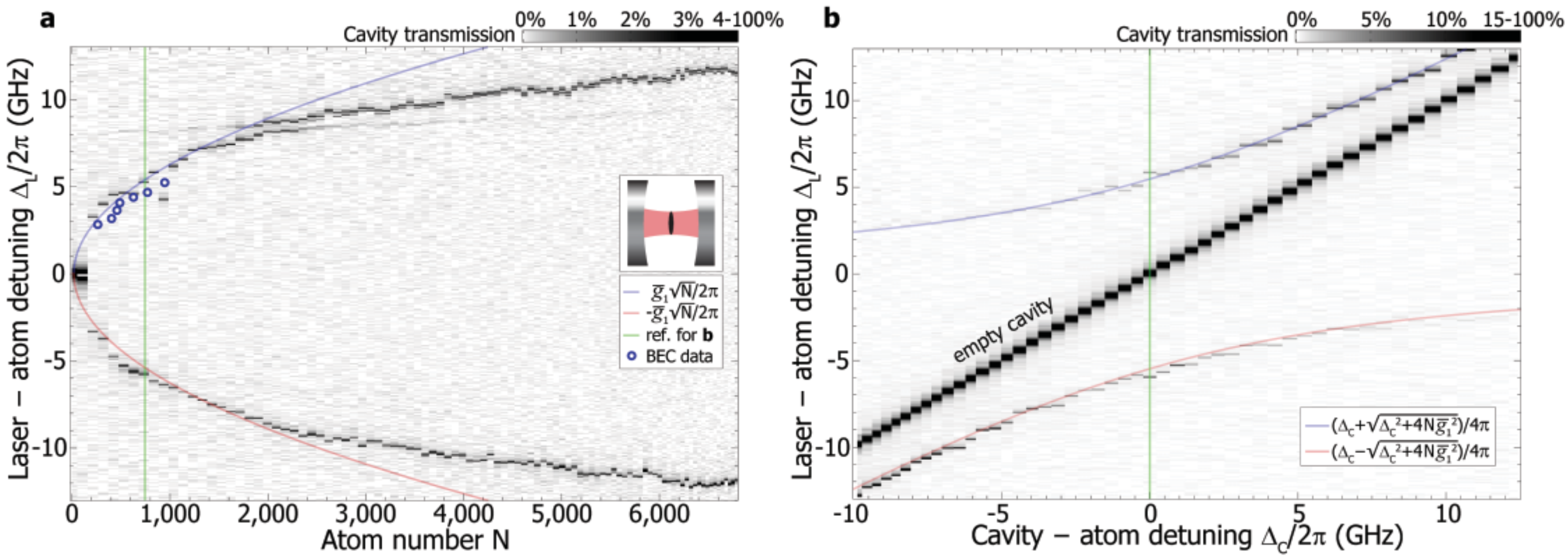

Figure 3: **Map of the energies of the dressed states**. **a**, $N$ is varied for $\Delta_C$=0. Greyscale spectra were taken on noncondensed atoms, circles were measured with BECs. The blue and red curves are the expected resonances at $\Delta_L = \pm\sqrt{N}\bar{g}_1$, where $\bar{g}_1$ is fitted to the noncondensed data for $N$<1000. The pictogram shows the orientation of the trap. **b**, $\Delta_C$ is varied for constant $N\approx750$; the empty cavity transmission ($N$=0) is also recorded and superimposed for reference. The green lines indicate experimental runs in **a** and **b** with common parameters $N$=750, $\Delta_C$=0. Probe intensity: $n_{res}\approx1.7\times10^{-1}$ (**a**) and $n_{res}\approx1.5\times10^{-1}$ (**b**), leading to $n\approx2.9\times10^{-2}$ (**a**) and $n\approx5.5\times10^{-2}$ (**b**) with atoms, at transmission peak.

A crucial feature of the collective coupling is its scaling with atom number. Early strong-coupling cQED experiments measured this in an atomic-beam apparatus with fluctuations both in $N$ and in the atomic spatial positions[25]. In our experiment, a BEC or a strongly confined cold thermal cloud minimize position fluctuations and $N$ remains fixed (as long as the interaction time is short enough to induce no significant losses, which is fulfilled here). We vary $N$ by forced evaporation, which produces

BECs for $N<3000$. The transmission is measured at $\Delta_C=0$ while sweeping the probe laser, $\Delta_L=2\pi\times(0\ldots\pm13)$GHz. $N$ is determined independently by absorption imaging (which underestimates the number of atoms, see Methods). We can either perform the measurement after transferring the atoms into a combined trap (same parameters as Fig. 2b), or directly in the magnetic trap, without optical lattice. The combined trap destroys BECs but increases the coupling when loading large thermal clouds; BECs confined in the magnetic trap alone remain intact after the measurement. The small, but measurable difference shown in figure 3a between the BEC (circles) and thermal cloud (greyscale) spectra is in accord with what is expected from the different density distributions (see Methods). The vacuum Rabi splitting reaches $2g_N=2\pi\times24$GHz for $N\approx7000$, corresponding to $C_N=4.4\times10^5$. For $N<1000$ the dressed state frequencies $E_\pm/\hbar$ have the expected $\pm\sqrt{N}\bar{g}_1$ dependence, with $\bar{g}_1\approx2\pi\times200\,\mathrm{MHz}$. The slower increase of the coupling for higher $N$ in the thermal cloud spectrum is due to the growing size of the sample. The anticrossing for $N\approx2000$ can be understood qualitatively on the reasonable assumption that a few atoms are in the $|F=1\rangle$ ground state. Atoms in $|F=2\rangle$ have transitions to the upper and lower dressed states at frequencies $\omega_A\pm g_N$ whereas $|F=1\rangle$ atoms have a transition at $\omega_A+\Delta_{HFS}$, $\Delta_{HFS}=2\pi\times6.8$GHz being the ground state hyperfine splitting of $^{87}$Rb. An anticrossing appears when the transition frequencies coincide at $g_N=\Delta_{HFS}$. However, this simple model, as well as the model used in ref. 12, cannot explain why the anticrossing occurs at larger detuning $\Delta_L\approx2\pi\times8.5$GHz (see Supplementary Notes). A complete understanding of this effect requires further investigation.

Figure 3b shows a measurement of the complete dressed-state spectrum (eq. 1). Conditions are as in Fig. 3a, but now $N\approx750$ is held constant and $\Delta_C$ is varied. Again, the observed resonances are in good agreement with the expected eigenfrequencies $E_\pm/\hbar$ plotted for $\bar{g}_1=2\pi\times200\,\mathrm{MHz}$. For large cavity detunings $|\Delta_C|>>g_N$ the dressed states of the system evolve toward the uncoupled states $|g{:}N\text{-}1;e{:}1;n=0\rangle$, where a single excitation is shared by all the atoms, and $|g{:}N;e{:}0;n=1\rangle$ for which the cavity contains a photon.

Finally, we measure the heating of a BEC caused by the intracavity field. To minimize technical heating, we use a purely magnetic trap ($\nu_x$=230Hz, $\nu_{y,z}$=2.0kHz). The BEC contains $N\approx800$ atoms and we use $n_{res}\approx3.9\times10^{-3}$, $\Delta_L=\Delta_C=0$, and an interaction time of 10ms. The BEC remains intact after the interaction with no measurable loss or heating if we maximize the coupling by positioning it on the cavity axis. We can vary the coupling by positioning the BEC at different heights $z_a$ relative to the centre of the mode. Figure 4 shows that the cavity transmission quickly drops to 0 as the coupling increases, because the strongly coupled system is no longer resonant for $\Delta_L=\Delta_C=0$. (Note that the probe light is then mostly reflected from the nonresonant cavity, not scattered by the atoms.) The heating rate of the condensate is measured using TOF imaging after the interaction (Fig. 4). It exhibits two peaks corresponding to measurements where $g_N$ is high enough for the atoms to be excited, but still low enough to allow a non-zero intracavity field. In these regions the BEC is destroyed after the interaction, whereas it is left unaffected for $|z_a|>18$μm and $|z_a|<5$μm, corresponding to $g_N<2\pi\times 10^{-3}$MHz and $g_N>2\pi\times 10^{3}$MHz. $g_N$ is calculated from $z_a$ and the measured $N$, averaging over the standing wave. The observed heating rate can be accounted for with a simple momentum-diffusion model[26,27] (see Supplementary Notes). This model predicts a peak heating rate about twice the observed value of ~400 cycles/s per atom, which is a satisfactory agreement given the uncertainty of our atom number calibration. For maximum coupling, the model yields a heating rate of $3\times 10^{-2}$ cycles/s per atom, which means that the whole BEC scatters an average of 0.24 photons during the interaction time.

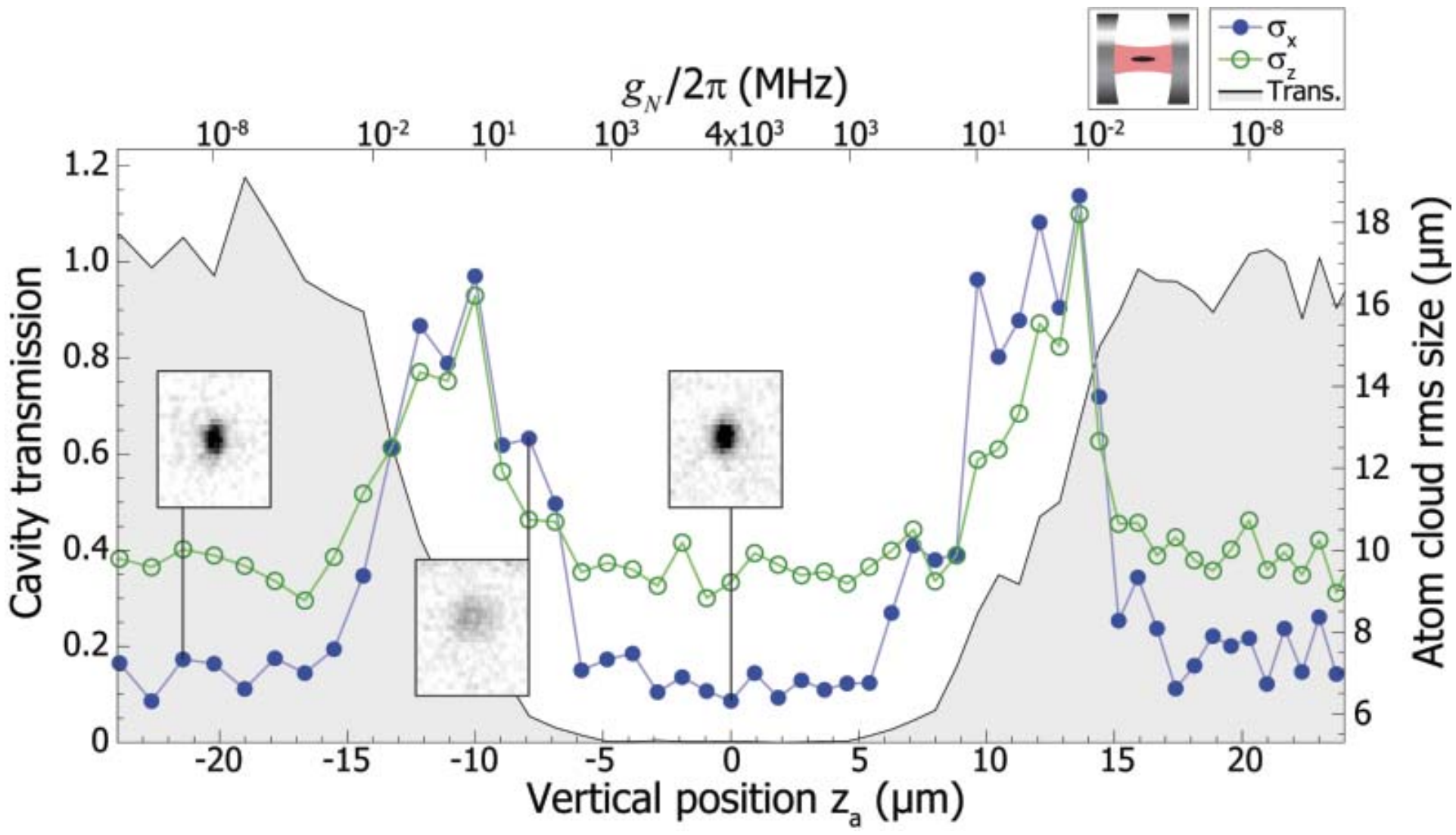

Figure 4: **Cavity-induced heating of the BEC at different transverse positions for $\Delta_L=\Delta_C=0$.** The pictogram shows the orientation of the trap. $g_N$ is varied by positioning the BEC at different heights $z_a$ relative to the cavity axis. The BEC causes a drop of the cavity transmission (line with shading under) when it is sufficiently coupled to the mode. The rms size along $x$ (dots) and $z$ (circles) of the atom cloud after a 2.8ms time-of-flight shows two heating peaks when the BEC is positioned such that $g_N \approx 2\pi\times 10$MHz, and no detectable heating for large coupling ($g_N > 2\pi\times 1$GHz). The insets show example TOF images.

In the experiments reported here, the small size of Bose-Einstein condensates was essential to produce an atomic ensemble with an extremely large, well-controlled, and homogeneous coupling rate. While the quantum degeneracy appears to leave no trace in the interaction with photons, our system is well-suited to directly study BEC atom statistics[8]. The cavity should allow QND measurement of the BEC atom number[28] and can be used as a single-atom detector with high quantum efficiency. Such a detector is sensitive to the internal atomic state and therefore highly suitable as a qubit detector. In the regime of atomic ensembles, compared to quantum light-matter interface experiments using noncondensed atoms in a cavity[16], the BEC additionally offers collisional interaction between atoms which can reach large, well-defined values and can be used as a resource[29]. Proposals exist, for example, to use Raman transitions for transferring a small, exactly known number $N_e$ of BEC atoms into a different internal state[30]. With the addition of a transverse laser beam, the cavity could be used to convert such state into a $N_e$-photon Fock state. Another, more technical advantage is the inherent fibre coupling of our cavities. We expect all of these properties to become important in future experiments.

**Methods Summary**

Our setup features two FFP cavities mounted on an atom chip with 150µm distance between their optical axes and the chip surface (Fig. 1). Both cavities are tunable independently over a full spectral range with piezoelectric actuators. The FFP1 cavity used in the experiments has length $d$=38.6µm, waist radius $w_0$=3.9µm, finesse $\mathcal{F}$=37000 and field decay rate $\kappa = \frac{\pi c}{2\mathcal{F}d} = 2\pi \times 53\text{MHz}$. The calculated maximum single-atom coupling rate is $g_0=2\pi\times 215\text{MHz}$, yielding a resonant saturation photon number $n_0=\gamma^2/(2g_0^2)\approx 1\times 10^{-4}$. The transmitted probe and lattice beams are separated after the output fibre, and the probe beam is detected with a photon-counting avalanche photodiode. The probe laser can be swept continuously over a range $\Delta_L=\omega_L-\omega_A=\pm 2\pi\times 15\text{GHz}$, where $\omega_A$ is the frequency of the $5S_{1/2}|F=2\rangle\rightarrow 5P_{3/2}|F'=3\rangle$ transition of $^{87}Rb$ at $\lambda_A$=780.2nm. BEC preparation is similar to our previous work. Absorption imaging inside and below the cavity is used for temperature and atom number measurements. Several effects lead to an underestimation of $N$ in these images; atom numbers extracted from the cQED measurements by using calculated $\bar{g}_1$ values are systematically higher by about a factor of 2.

## Methods

**Fibre Fabry-Perot cavity.** For a two-level atom in a symmetric Fabry-Perot cavity (mirror distance $d$ and radius of curvature $r$), $g_0 = \sqrt{\frac{3\lambda^2 c\gamma}{\pi^2 w_0^2 d}} = \sqrt{\frac{6\lambda c\gamma}{\pi d\sqrt{2dr - d^2}}}$ and $C_0 = \frac{3\lambda^2 \mathcal{F}}{\pi^3 w_0^2} = \frac{6\lambda \mathcal{F}}{\pi^2 \sqrt{2dr - d^2}}$, where $w_0$ is the mode waist radius (which depends on the mirror distance and curvature) and $\mathcal{F} \approx \frac{\pi}{T+L}$ is the cavity finesse, which depends solely on the intensity transmission $T$ and loss $L$ of each mirror, but not on geometry. The second expression for $C_0$ shows that short cavities with strongly curved mirrors lead to high cooperativity. In our FFP cavity[10], the concave mirror surfaces are realized on the cleaved surfaces of two optical fibre tips facing each other (Fig. 1b,c). With this type of cavity, the optical axis can approach the chip surface to half a fibre diameter, and maximum coupling is achieved by placing the atoms in the gap between the mirrors, so that their distance from any material surface remains in the many-micrometre range, where long coherent trapping times have been demonstrated. In the second-generation FFP fabrication method employed here, a laser surface machining process is used to shape the mirror surfaces, which has allowed us to achieve the finesse of $\mathcal{F}$ =37000 for the cavity used here, and should ultimately enable $\mathcal{F}$ =150000 (for mirrors where transmission equals total loss), based on the measured surface roughness. Additionally, this method can produce very small radii of curvature. The FFP1 cavity used here has an asymmetric geometry with radii $r_1$=450µm and $r_2$=150µm (measured by atomic force microscopy), and uses a single-mode (SM) fibre on the input side, while a multi-mode (MM) output fibre on the output side assures high outcoupling efficiency (Fig. 1c). There is a splitting between modes of orthogonal linear polarization, which is 540MHz between the TEM00 modes in our cavity. This allows us to adjust probe beam polarization to one particular linear polarization. However, due to the direct fibre coupling, we are currently unable to determine the axis of this polarization in the $yz$ plane. The cavity length $d$=38.6µm (longitudinal mode number $n$=99) is confirmed by two-frequency transmission measurements, and the waist radius $w_0$=3.9µm is inferred from this $d$ and the mirror curvatures. These parameters lead to the $g_0$ value quoted in the main text. With a Rayleigh length of about 60µm, the cavity mode is quasi-cylindrical: the variation of the beam radius $w(x)$ is 12%. (In the fit of Fig. 2b, this variation is taken into account.) The resonance linewidth $2\kappa$ is measured with two frequency-stabilized lasers. The transmission of each cavity mirror is $T$=31±2ppm as determined from a reference substrate coated in the same batch as the fibres. From this value and the measured finesse, we infer per-mirror intensity losses $L$=56ppm. The measured intensity transmission from before the input of the SM fibre to after the output of the MM fibre is 0.094 for resonant cavity. Comparison to the calculated transmission of the cavity $\left(\frac{T}{T+L}\right)^2 = 0.126$ indicates that the combined losses of coupling into the SM fibre, mode-matching into the cavity, and from the cavity to the SM fibre are 0.253, a low value. The cavity length is actively stabilized to compensate for thermal drifts of ~1500 linewidths. The locking scheme uses a correction signal from FFP2 which is locked on a resonance and subjected to the same thermal perturbations as FFP1; residual drifts of a few linewidths are corrected using an error signal derived from the 830nm lattice beam.

**Atom chip and BEC production.** The cavity subassembly is glued onto an atom chip, which forms the top wall of a commercial glass cell[31]. Sealed fibre feedthroughs are formed simply by two slits machined into opposite walls of the cell, which are filled with vacuum-compatible epoxy glue once the fibres are in place. Base pressure in the cell is $3\times10^{-10}$hPa, comparable to the pressure in similar cells without cavities. As in our previous work[32], we use a mirror-MOT to precool $^{87}$Rb atoms. To avoid obscuring of MOT beams by the cavity subassembly, the horizontal MOT beams are parallel to the *x* axis and there is a distance of 11mm along *y* between the MOT (position A in Fig. 1a) and cavity centres. After optical pumping to the $|F=2,m_F=2\rangle$ state and initial magnetic trapping near the MOT location and magnetic transport (which combines a wire guide[11] and an external quadrupole field), the atoms are further cooled by forced radiofrequency evaporation in a "dimple" trap[32] (formed by a wire cross with currents 3A and 300mA) between the chip surface and the cavity mode at position B (Fig. 1b); the BEC is produced 17µm above the mode axis. Alternatively, in some experiments, we use surface evaporation[33, 34] near one of the fibre endfaces, as we found that this increases atom number stability for small condensates. The trap geometry is prolate (cigar-shaped); by ramping wire currents we can align its long axis parallel or perpendicular to the cavity axis[35], as required in a particular experiment. Absorption imaging inside and above the cavity is possible with a probe beam in the *yz* plane, which subtends an angle of 30 degrees with the chip surface and is reflected by the dielectric coating on the chip before its passage through the cavity (Fig. 1d). This reflection limits the achievable purity of the desired circular polarization; furthermore, camera noise in conjunction with the magnification of about 4 forces us to use a relatively high imaging beam intensity that causes some saturation. We have not attempted to correct these effects, all of which lead to an underestimation of *N*. Atom numbers extracted from the cQED measurements by using calculated $\bar{g}_1$ values are systematically higher by about a factor of 2.

**BEC coupling strength.** In Fig. 3a the circles show the coupling strength $g_N$ for BECs held in a magnetic trap ($\nu_{x,z}$=2.7kHz, $\nu_y$=230Hz, bias field $B_y\approx$1G), with atom numbers N ranging from 260 to 950. $g_N$ is reduced by a factor of about 0.86 (mean value over the 7 measurements) with respect to the values measured for thermal samples in the combined trap (same trap parameters as in Fig. 2b). The expected reduction in $g_N$ is calculated using the density distributions of the BECs in the 1d-3d crossover regime[24] and the density distribution of thermal clouds at 4.4μK, as obtained from the fit in Fig. 2b. This assumes that the final temperature of the sample is independent of *N* when loading BECs. We find a reduction by a factor 0.83 (mean value), in good agreement with the measured value. For these measurements a very weak probe beam is used, $n_{\rm res}=6.3\times10^{-4}$, in order to leave the condensate unaffected. For each of the 7 data points we average over 4 identical experimental runs to obtain a clear transmission signal.

**Acknowledgements** We thank J. Hare and F. Orucevic for their support in producing the fibre mirror surfaces, and F. Gerbier for the calculation of condensate size in the crossover regime. We gratefully acknowledge fruitful discussions with Y. Castin and J. Dalibard about atom-light interaction in BECs, as well as with T. W. Hänsch, I. Cirac, P. Treutlein, and R. Long. This work was supported by a European Young Investigator Award (EURYI), a Chaire d'Excellence of the French Ministry for Research, and by the E.U. ("Atom Chips" Research Training Network and "SCALA" Integrated Programme). The Atom Chip team at Laboratoire Kastler Brossel is part of the Institut Francilien de Recherche sur les Atomes Froids (IFRAF).

Correspondence and requests for materials should be addressed to J. R. (jakob.reichel@ens.fr).

# Supplementary Notes

## 1 Conditions for collective interaction

This section summarizes some facts about collective interaction of $N$ two-level atoms with a single mode of the radiation field in the strong-coupling regime.

Let us first consider $N$ atoms at fixed positions, all identically coupled to a single mode of the radiation field [S1] with a coupling strength $g_1$. We call $|\mathrm{g}_i\rangle$ and $|\mathrm{e}_i\rangle$ the ground and excited internal states of the $i$th atom. If initially all atoms are in the ground state $|\Psi_0\rangle = |\mathrm{g}\ldots\mathrm{g}\rangle$ and then the system is weakly excited, the first excited state must reflect the symmetry of the situation and therefore is [S2]

$$|\Psi_1\rangle = \frac{1}{\sqrt{N}}\left(|\mathrm{e},\mathrm{g}\ldots\mathrm{g}\rangle + |\mathrm{g},\mathrm{e},\mathrm{g}\ldots\mathrm{g}\rangle + \ldots + |\mathrm{g}\ldots\mathrm{g},\mathrm{e}\rangle\right)\,. \tag{1}$$

The coupling term in the Hamiltonian, setting $\hbar = 1$, is

$$\widehat{V} = g_1\left(\sum_i \hat{a}|\mathrm{e}_i\rangle\langle\mathrm{g}_i|\right) + \mathrm{h.c.}\,, \tag{2}$$

where $\hat{a}$ annihilates one photon in the mode. The matrix element $g_N$ between the atomic ground state with one photon in the field mode, $|1\rangle\otimes|\Psi_0\rangle$, and the atomic first excited state with no photon, $|0\rangle\otimes|\Psi_1\rangle$, is therefore

$$g_N = \langle 0|\otimes\langle\Psi_1|\,\widehat{V}\,|\Psi_0\rangle\otimes|1\rangle = \sqrt{N}g_1\,. \tag{3}$$

The atomic ensemble thus behaves as a single "superatom", with a collective enhancement $\sqrt{N}$ in the coupling strength.

No assumption about the particle statistics was made to obtain this result. However identical coupling was assumed, which in a real experiment is strictly true only for a Bose-Einstein condensate (BEC), where all atoms share the same spatial wavefunction. In the following we show that even for nonidentical couplings the atomic ensemble interacting with the cavity field can be described as a two-level system, and that the $\sqrt{N}$ scaling of the collective coupling strength still holds. The main result is that $g_N = \sqrt{N}\overline{g}_1$ with $\overline{g}_1^{\,2} = \int \frac{\rho(\mathbf{r})}{N}|g_1(\mathbf{r})|^2\mathrm{d}\mathbf{r}$, where $g_1(\mathbf{r})$ is the position-dependent single-atom coupling strength to the field mode and $\rho(\mathbf{r})$ is the atomic density distribution. Additionally, the reasoning below does not make the approximation of point-like atoms, so that it applies to laser-cooled atoms and BECs for which the size of the atomic wavefunction can easily exceed the optical wavelength (note that the demonstration would be much simpler using the point-like approximation). The case of a *single* atom with quantized motion was considered by Vernooy and Kimble [S3], who found complex dynamics with collapses and revivals of the Rabi oscillations between the ground and excited states of the atom.

The complete (neglecting damping) $N$ atoms - field Hamiltonian in a cavity can be written

$$\widehat{H} = \underbrace{\omega_\mathrm{C}\,\hat{a}^\dagger\hat{a} + \omega_\mathrm{A}\sum_{i=1}^{N}\hat{\sigma}_i^\dagger\hat{\sigma}_i}_{\widehat{H}_0} + \underbrace{\sum_{i=1}^{N}\left(g_1(\hat{\mathbf{r}}_i)\,\hat{a}\hat{\sigma}_i^\dagger + g_1^*(\hat{\mathbf{r}}_i)\,\hat{a}^\dagger\hat{\sigma}_i\right)}_{\widehat{V}}\,, \tag{4}$$

where $\hat{\sigma}_i$ is the atomic lowering operator $\hat{\sigma}_i = |\mathrm{g}_i\rangle\langle\mathrm{e}_i|$ and $\hat{a}$ the cavity photon destruction operator. The kinetic energy of the atoms is neglected, which is valid as long as the broad line condition $E_\mathrm{rec} \ll \hbar\,\Gamma$ applies, meaning that a few recoil kicks will not drive the atoms out of resonance (see for example [S4]).

Suppose that the atom-cavity compound system is initially is the stationary state $|G\rangle \equiv |0;\mathrm{g}\ldots\mathrm{g};\Phi\rangle$: no photon is in the cavity, all the atoms are in the internal ground state $|\mathrm{g}\rangle$ and their positions are described by a $N$-particle wavefunction $|\Phi\rangle$. When the cavity is pumped, the system can undergo

transitions to the first excited state $|E_1\rangle \equiv |1; \mathrm{g}\ldots\mathrm{g};\Phi\rangle$. This level is coupled via $\widehat{V}$ to a second excited state $|E_2\rangle$ defined by:

$$|E_2\rangle = \widehat{V}|E_1\rangle = |0\rangle \otimes \sum_i |\mathrm{g}\ldots \underbrace{\mathrm{e}}_{i^{th}term} \ldots\mathrm{g}\rangle \otimes g_1(\hat{\mathbf{r}}_i)|\Phi\rangle \tag{5}$$

up to a normalization constant. To confirm that the atomic ensemble interacting with the cavity field behaves like a two-level system, one has to show that the action of $\widehat{V}$ on $|E_2\rangle$ is mainly a coupling to $|E_1\rangle$. Therefore one has to compute

$$\widehat{V}|E_2\rangle = |1;\mathrm{g}\ldots\mathrm{g}\rangle \otimes \sum_i |g_1(\hat{\mathbf{r}}_i)|^2|\Phi\rangle\,. \tag{6}$$

When the two-particle position correlations are small, as in a classical gas or in a Bose-Einstein condensate, one can show that the final position wavefunction differs only slightly from the initial one:

$$\sum_i |g_1(\hat{\mathbf{r}}_i)|^2|\Phi\rangle = \left(\sum_i \left\langle |g_1(\hat{\mathbf{r}}_i)|^2\right\rangle_\Phi\right)\left(|\Phi\rangle + \mathcal{O}(\frac{1}{\sqrt{N}})\right) = g_N^2|\Phi\rangle + \mathcal{O}(\sqrt{N})\,, \tag{7}$$

$< \cdot >_\Phi$ standing for the quantum average (expectation value) in the state $|\Phi\rangle$. This would not be true for a quantum state presenting strong correlations, such as a Schrödinger cat, for which the position of the atoms would become highly entangled with the internal state.

Finally, the coupling $\widehat{V}$ can be rewritten as the sum of a two-level coupling $\widehat{V}_{JC}$, and a many-level coupling term $\widehat{V}'$ of smaller magnitude, which corresponds to cloud heating and spreading on a longer timescale due to interaction with the cavity field. The two-level coupling can be cast in the familiar Jaynes-Cummings form $\widehat{V}_{JC} = g_N(\hat{a}\hat{\sigma}^\dagger + \hat{a}^\dagger\hat{\sigma})$ with an effective, scalar coupling strength $g_N$ scaling as $\sqrt{N}$, as in the case of identically coupled point-like atoms (3):

$$g_N^2 = \sum_{i=1}^N \left\langle |g_1(\hat{\mathbf{r}}_i)|^2\right\rangle_\Phi = N\left\langle |g_1(\hat{\mathbf{r}}_i)|^2\right\rangle_{\Phi,i} = \int \langle\hat{\rho}(\mathbf{r})\rangle_\Phi\,|g_1(\mathbf{r})|^2\mathrm{d}\mathbf{r}\,, \tag{8}$$

and hence

$$g_N = \sqrt{N}\overline{g}_1\,,\ \text{with}\ \overline{g}_1^2 = \int \frac{\rho(\mathbf{r})}{N}|g_1(\mathbf{r})|^2\mathrm{d}\mathbf{r}\,. \tag{9}$$

## 2 Multilevel coupling model

In our experiments, $^{87}$Rb atoms are trapped in the $|F=2\rangle$ hyperfine level of the $5S_{1/2}$ ground state. We note $\Delta_L = \omega_L - \omega_A$ the detuning of the probe beam relative to the $5S_{1/2}\,|F=2\rangle \to 5P_{3/2}\,|F'=3\rangle$ line of the $D_2$ transition. When $|\Delta_L|$ exceeds the $\sim$ 500 MHz hyperfine splitting of the $5P_{3/2}$ excited state, one can treat the $5S_{1/2}\,|F=2\rangle \to 5P_{3/2}\,|F'=1,2,3\rangle$ transitions as a whole, which allows a two-level approach for the atoms. However, at the very large atoms-cavity coupling strengths reached in the experiment, additional atomic levels may play a role in the coupled system. In particular, when the coupling $g_N$ is of the order of the hyperfine splitting of the $5S_{1/2}$ ground state $\Delta_{HFS} = (E_{F=2} - E_{F=1})/\hbar \approx 2\pi \times 6.8\,\text{GHz}$, the transitions $|F=1\rangle \to 5P_{3/2}$ and $|F=2\rangle \to |\text{upper dressed state}\rangle$ may become simultaneously resonant with the probe field. As we will see, this leads to new features in the spectrum provided that the $|F=1\rangle$ state is populated, even by a small fraction of atoms.

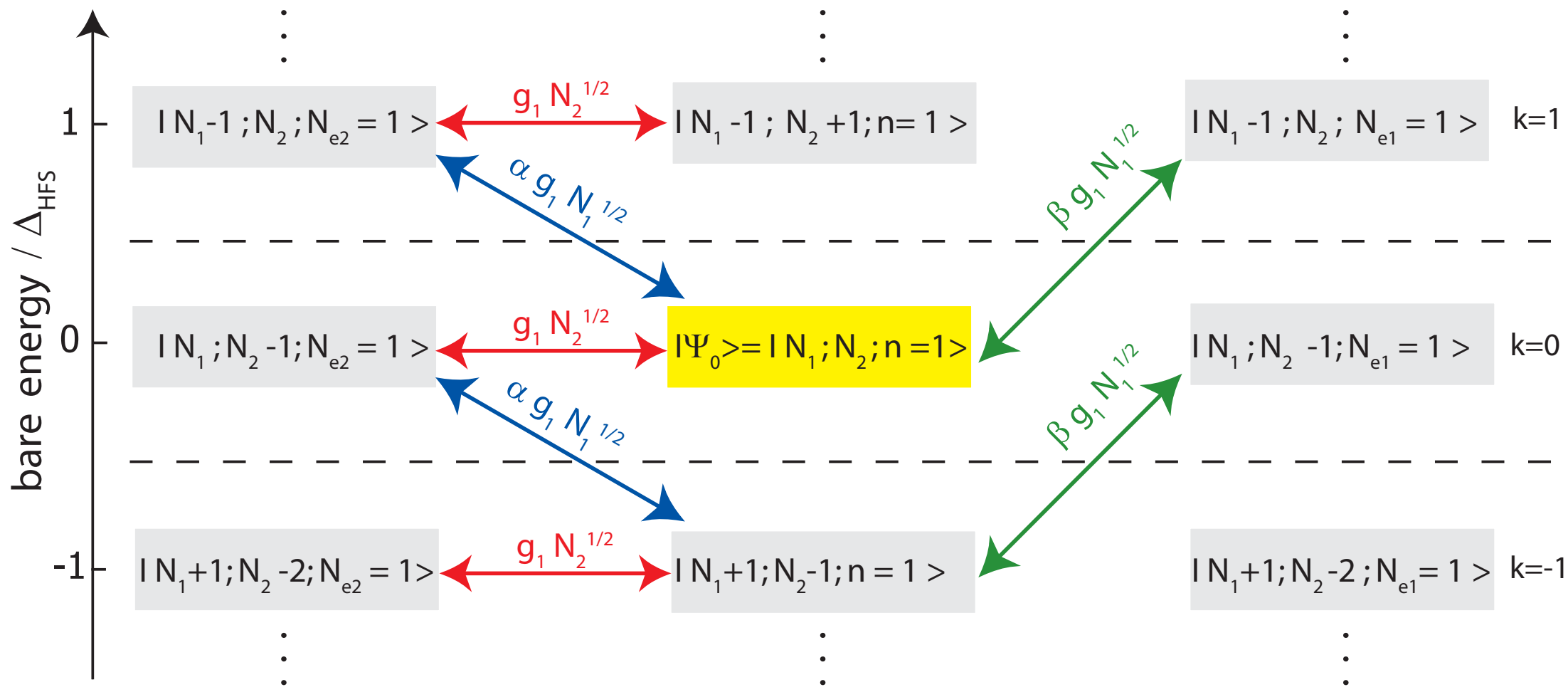

**Supplementary figure** 1: Outline of the first-excited states involved in the calculation of the cavity transmission spectrum. Left and right column states are "atom-like" (they contain one excited atom), central column states are "cavity-like" (they contain one photon). The vertical axis is the bare energy, in units of the ground state hyperfine splitting $\Delta_{HFS}$. The initial state $|\Psi_0\rangle$ is in the centre. Horizontal and oblique arrows indicate the coupling strengths. The dashed lines separate the $k$ subspaces; we restrict the graph to $k=-1,0,1$ for clarity. The cavity is tuned on the $5S_{1/2}\,|F=2\rangle \to 5P_{3/2}$ transition, as in the experiment of Fig. 3a (main text).

A simple model taking into account this new transition is a four-level atom, with ground states $|g_{1,2}\rangle \equiv 5S_{1/2}\,|F=1,2\rangle$ of energies $-\hbar\Delta_{HFS}$ and 0, and excited states $|e_{1,2}\rangle$ of energy $\hbar\omega_A$. The cavity coupling can be developed on this basis (in the rotating wave approximation):

$$\widehat{V}_1 = \widehat{\mathbf{D}} \cdot \widehat{\mathbf{E}} = \hbar g_1\, \hat{a}\Big(|e_1\rangle\langle g_2| + \alpha|e_1\rangle\langle g_1| + \beta|e_2\rangle\langle g_1|\Big) + \text{h.c.} \tag{10}$$

in which the parameters $\alpha, \beta \lesssim 1$ depend on the polarization of the cavity field. A generic state for the $N$-atoms + cavity system is described by the number of atoms in each state and the number of photons in the cavity: $|\Psi\rangle = |N_{g_1}; N_{g_2}; N_{e_1}; N_{e_2}; n\rangle$. The initial state is $|N_1; N_2; 0; 0; 0\rangle$, where it is hypothesized that a relatively small number of atoms $N_1 \ll N_2$ are in the $|F=1\rangle$ state. When the cavity is weakly pumped with the probe laser, the system is excited to the "cavity-like" state $|\Psi_0\rangle \equiv |N_1; N_2; 0; 0; 1\rangle$. This level is coupled to many other states with the same number of excitations $M = N_{e_1} + N_{e_2} + n = 1$, which we write $|N_1 - k; N_2 + k - N_e; N_{e_1}; N_{e_2}; n\rangle$, with $N_e = N_{e_1} + N_{e_2}$ and $k$ an integer (Supplementary figure 1). For growing $|k|$ values, these levels of energy $\sim k \times \hbar\Delta_{HFS}$ get progressively out of resonance with $|\Psi_0\rangle$. We thus restrict $|k|$ to small values, $k = -3 \ldots 3$, and diagonalize the coupling in this $3 \times 7$ - dimensional subspace. Each eigenenergy $E$ is related to a peak in the frequency spectrum at a detuning $\delta \equiv (E - E_{\Psi_0})/\hbar = \omega_L - \omega_C$, with a weight $\sim |\langle\Psi|\Psi_0\rangle|^2$.

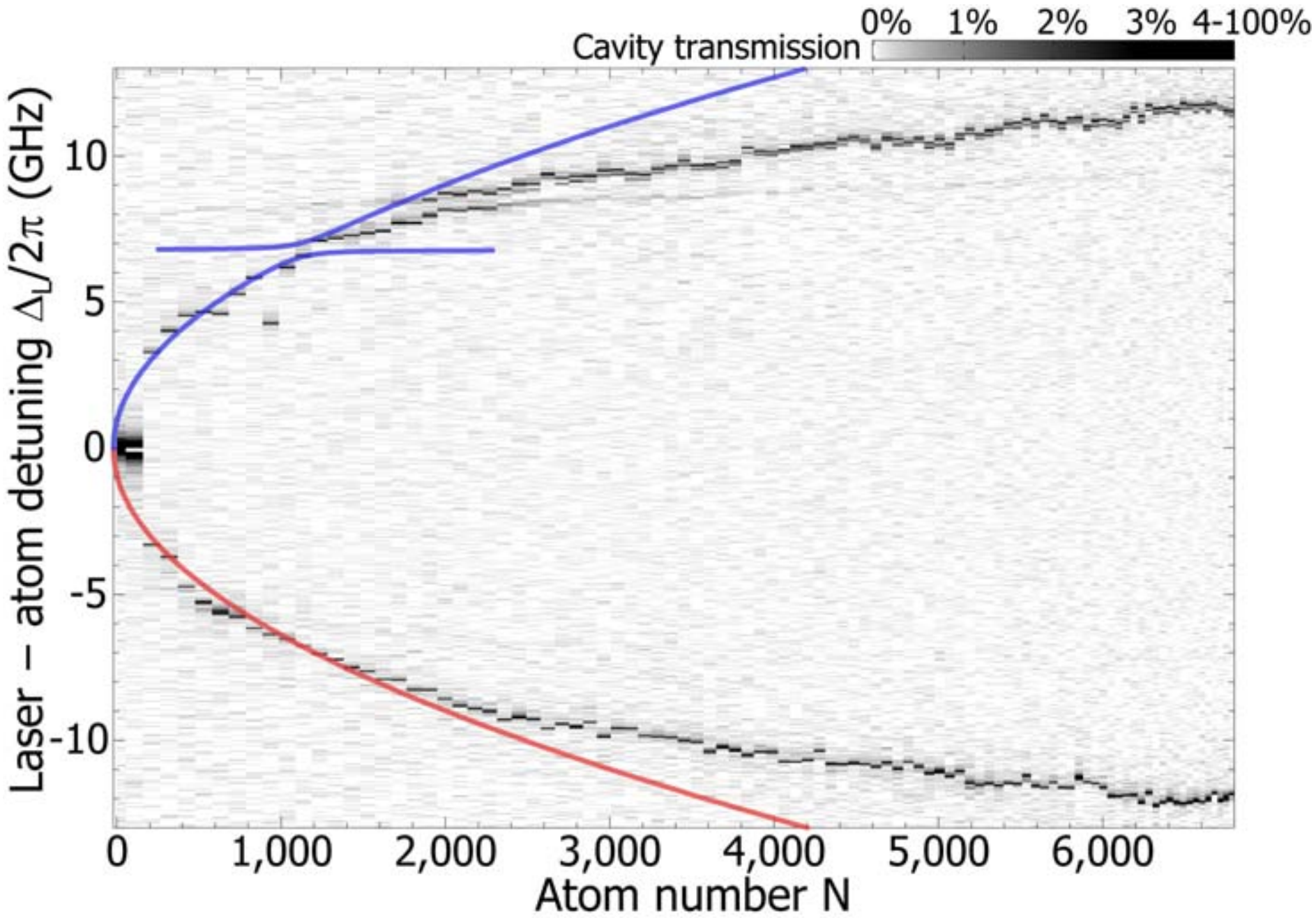

**Supplementary figure** 2: Eigenenergies computed using the simple multilevel model with $\Delta_C = 0$, $\alpha = 0$, $\beta = 1$, $N_1/N_2 = 0.25\%$, superimposed on the experimental data of Fig. 3a (main text).

When the cavity is tuned on the $5S_{1/2}\,|F=2\rangle \rightarrow 5P_{3/2}$ transition ($\Delta_C = 0$), the model predicts an anticrossing in the positive part of the spectrum ($\delta > 0$) when the double resonance condition $g_1\sqrt{N_2} = \Delta_{HFS}$ is fulfilled. It is located at $\delta = \Delta_{HFS}$ and has a frequency gap $\sim \beta g_1\sqrt{N_1}$. We adjust the parameters of the model to reproduce the frequency gap $\sim 500\,\text{MHz}$ of the experimental data (Fig. 3a in the main text). Since the polarization of the cavity field is not precisely known, $\beta$ is unknown and set to 1 for simplicity. The only magnetically trapped state of $5S_{1/2}\,|F=1\rangle$ is $|F=1, m_F=-1\rangle$, hence we assume that $|F=1\rangle$ atoms should be this Zeeman substate. As $\alpha \propto \langle g_2|\hat{V}_1^2|g_1\rangle$, $\alpha$ is expected to be zero since there is no 2-photon transition between states $|F=2, m_F=2\rangle$ and $|F=1, m_F=-1\rangle$. A finite value for $\alpha$ would lead to additional anticrossings in the spectrum, located at detunings $\delta = \pm\Delta_{HFS}/2$, which we do not observe in this experiment. The remaining parameter is the $|F=1\rangle$ population fraction $N_1/N_2$ – supposed constant regardless of the total number of atoms – which we fit to the data, finding $N_1/N_2 = 0.25\%$. The corresponding resonance curves are shown in Supplementary figure 2. The anticrossing predicted by the model does not occur at the detuning at which it is observed experimentally ($\delta_{exp} \approx 8.5\,\text{GHz}$). Note also that the asymptotes in the experimental spectrum clearly have nonzero slope, whereas the model predicts horizontal asymptotes. In a more refined model we have taken into account the optical pumping of the atoms arising during the excitation of the coupled system, but found that this effect remains negligible.

An anticrossing related to the hyperfine structure of the ground state is observed by Brennecke *et al.* [S5]. The authors develop a model similar to ours, but including also the complete structure of the $D_2$ transition, the presence of two light polarizations in the cavity, and a "Lamb-shift" effect due to higher order transverse modes of the cavity. However, as we have checked, none of these effects can account for the position of the anticrossing or the non-zero slope of the asymptotes present in our experimental data. In particular, the first higher order transverse mode of our FFP1 cavity lies some 750 GHz above the $TEM_{00}$ mode, which is too far detuned to affect the measured spectrum.

Further theoretical and experimental investigations are required to explain the origin and characteristics of the observed anticrossing.

# 3 Heating rate model

In the experiments shown in Fig. 4 (main text), a magnetically trapped BEC interacts with the cavity field for $t_{\rm int} = 10\,$ms, with $\Delta_{\rm L} = \Delta_{\rm C} = 0$. The heating of the trapped cloud is measured using absorption imaging after a time-of-flight (TOF) expansion $t_{\rm TOF} = 2.8\,$ms.

We calculate the heating rate of the trapped cloud with a momentum diffusion model. Fischer *et al.* [S6] derive the diffusion for an untrapped atomic cloud in a cavity in the low saturation limit. In the resonant case ($\Delta_{\rm L} = \Delta_{\rm C} = 0$), the momentum diffusion coefficient for the atom $i$, defined by $\frac{\rm d}{{\rm d}t}(\langle \mathbf{p}_i^2\rangle - \langle \mathbf{p}_i\rangle^2) \equiv 2D_p^i$, is

$$D_p^i = \frac{n(\hbar k)^2 g_1(\mathbf{r}_i)^2}{\gamma} + \frac{n\hbar^2(\boldsymbol{\nabla} g_1(\mathbf{r}_i))^2}{\gamma}\,, \tag{11}$$

where $n$ is the intracavity photon number in the presence of the $N$ atoms and $g_1(\mathbf{r}_i)$ is the coupling strength for the atom $i$, $g_1(\mathbf{r}_i) = g_0\cos(kx_i)\exp(-r_{\perp i}^2/w^2)$.

The first term in (11) can be rewritten $(\hbar k)^2\Gamma_{\rm sp}^i/2$, where $\Gamma_{\rm sp}^i$ is the rate of spontaneous emission of the atom $i$. This part of the diffusion process is equivalent to a random walk with step length $\hbar k$, at a rate $\Gamma_{\rm sp}^i$, and corresponds to the emission of spontaneous photons in random, isotropically distributed directions. The second term is associated with the randomness of the direction of the absorbed photons. Neglecting the radial contribution in $\boldsymbol{\nabla} g_1$, the two terms have the same average value as soon as the cloud size is large compared to $\lambda/2$. This is the case in the experiments of Fig. 4 (main text), where the calculated diameter of the condensate along the cavity axis is $2R_x = 6.6\,\mu$m.

Expressing $n$ in terms of the number of photons $n_{\rm res}$ in the empty cavity with the same probe beam intensity, we write the average diffusion as a function of the collective cooperativity $C_N$ :

$$D_p = \frac{2n_{\rm res}(\hbar k)^2\kappa}{N}\frac{2C_N}{(1+2C_N)^2}\,. \tag{12}$$

The diffusion is maximal for $C_N = 0.5$. Given the small radial extension of the BEC ($2\sigma_{y,z} = 490\,{\rm nm} \ll w = 3.9\,\mu$m), we neglect the variation of the coupling in the transverse directions over the atomic sample: $g_1(x, r_{\perp i}) \equiv g_1(x, r_\perp)$. This yields a straightforward radial dependence for $C_N$, $C_N = \frac{1}{2}NC_0\exp(-2\,r_\perp^2/w^2)$ where $C_0 = g_0^2/2\kappa\gamma$ is the maximum single-atom cooperativity. Replacing $N$, $C_0$ and $w$ with their values in the experiment, the diffusion peaks are expected to occur when the condensate is positioned about $10\,\mu$m off-axis, which is consistent with our measurements.

After the interaction time $t_{\rm int}$, the probe field is shut off and the atoms are brought to the centre of the mode in $t_{\rm transp} = 22\,$ms. This provides identical TOF conditions regardless of the initial radial position $|z_a| \equiv r_\perp$ of the cloud, and allows the sample to thermalize. We calculate the cloud rms sizes $\sigma_{x,y,z}$ after TOF assuming that the energy imparted by the heating is equally redistributed among the 6 degrees of freedom of the trap. This leads to

$$\sigma_x^2 \approx \sigma_{x,\rm ref}^2 + \frac{1}{3}\frac{D_p\, t_{\rm int}\, t_{\rm TOF}^2}{M_{\rm Rb}^2}\,, \tag{13}$$

and to similar expressions for the $y$ and $z$ directions. $M_{\rm Rb}$ is the $^{87}$Rb mass and $\sigma_{x,\rm ref}$ is the cloud rms size after TOF when the probe beam is kept off. In (13) we neglect the contribution of the increase in the trapped cloud size, which is a fair approximation. Supplementary figure 3 compares the calculated $\sigma_x$ with the experimental data; $\sigma_{x,\rm ref}$ is set to its measured value $\sigma_{x,\rm ref} \approx 7\,\mu$m. The calculated diffusion coefficient $D_p^{\rm peak} = n_{\rm res}(\hbar k)^2\kappa/2N$ at the two maxima of heating leads to an almost isotropic cloud after TOF expansion – as is observed experimentally – with $\sigma^{\rm peak} = v_{\rm rec}\,t_{\rm TOF}\,(n_{\rm res}\,\kappa\,t_{\rm int}/6N)^{1/2}$, where $v_{\rm rec}$ is the recoil velocity. The numerical value $\sigma^{\rm peak} = 28\,\mu$m overestimates the experimental value $\sigma_{\rm exp}^{\rm peak} = 18\,\mu$m . There are at least two reasonable explanations for that: first, the measured value for the atom number $N$ can be inaccurate up to a factor 2, yielding a factor $\sqrt{2}$ for the cloud final size, consistent with the other experiments which also suggest that the atom number is underestimated; second, as we estimate below, the number of trapped atoms changes during the interaction with the cavity due to optical pumping, thereby changing the value of $C_N$. This second effect may also explain why the diffusion peaks are broader than expected.

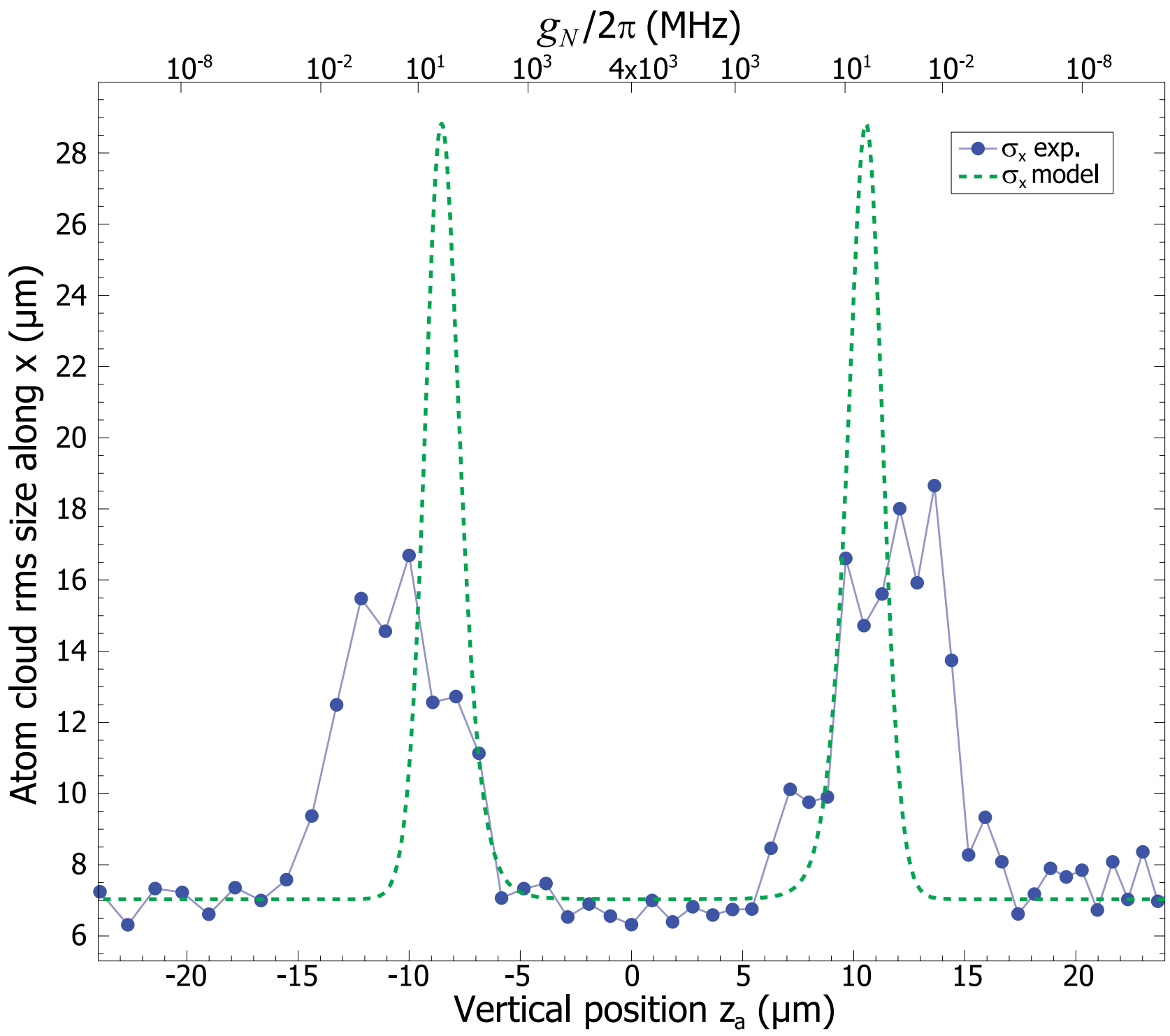

**Supplementary figure** 3: Calculated rms cloud size along $x$ after $t_{\mathrm{TOF}} = 2.8\,\mathrm{ms}$, compared to the experimental data shown in Fig. 4 (main text).

The model gives the number of spontaneous emissions per atom at the heating peaks $n_{\mathrm{sp}}^{\mathrm{peak}} = n_{\mathrm{res}}\,\kappa\,t_{\mathrm{int}}/2N \approx 8$, and allows to check the validity of the low-saturation assumption: $P_{\mathrm{exc}}^{\mathrm{peak}} = \Gamma_{\mathrm{sp}}^{\mathrm{peak}}/2\gamma = 2 \times 10^{-5} \ll 1$. Taking the direction $x$ of the magnetic field $\mathbf{B}_0$ at the bottom of the trap as quantization axis, a linear polarization in the cavity is an equal-weight combination of $\sigma^+$ and $\sigma^-$ polarizations. A crude calculation based on branching ratios shows that about 15% of the spontaneous emission events drive the magnetically trapped $|F = 2, m_F = 2\rangle$ atoms into the untrapped states $|F = 1, 2, m_F = 0\rangle$ (free falling) and $|F = 1, m_F = 1\rangle$ (expelled from the trap). This means that a fraction $\gtrsim 0.85^8 = 27\%$ of the atoms stay in the initial state after interaction, in rough agreement with the measurements (the number of detected atoms drops by about 50% at the heating peaks).